# Revisiting Fully Homomorphic Encryption Schemes

Nimish Jain     Aswani Kumar Cherukuri*

*School of Information Technology and Engineering (SITE)*

*Vellore Institute of Technology, Vellore*

*[*cherukuri@acm.org*](mailto:cherukuri@acm.org)

## Abstract

Homomorphic encryption is a sophisticated encryption technique that allows computations on encrypted data to be done without the requirement for decryption. This trait makes homomorphic encryption appropriate for safe computation in scenarios involving sensitive data, such as cloud computing, medical data exchange, and financial transactions. The data is encrypted using a public key in homomorphic encryption, and the calculation is conducted on the encrypted data using an algorithm that retains the encryption. The computed result is then decrypted with a private key to acquire the final output. This abstract notion protects data while allowing complicated computations to be done on the encrypted data, resulting in a secure and efficient approach to analysing sensitive information. This article is intended to give a clear idea about the various fully Homomorphic Encryption Schemes present in the literature, as well as analysing and comparing the results of each of these schemes. Further we also provide applications and open source tools of homomorphic encryption schemes.

## 1. INTRODUCTION

The term "homomorphic" is derived from two Greek roots: "homo," which means "same," and "morph," which means "shape." The term homomorphism in mathematics refers to a structure-preserving map between two algebraic systems whose operations are the same or similar. The phrase "homomorphic encryption" refers to how this encryption approach allows computations to be conducted on encrypted data while preserving the data's structure, allowing the same computations to be performed on encrypted data as on unencrypted data. Homomorphic Encryption (HE) is a kind of encryption scheme that allows a third party (e.g., cloud, service provider) to perform certain computable functions on the encrypted data while preserving the features of the function and format of the encrypted data.

Encryption methods like RSA, AES, and DES are not homomorphic, meaning that they require the data to be decrypted before any computation can be performed. This makes it challenging to use these encryption methods in situations where data privacy is a critical concern, such as cloud computing and data analytics. In contrast, homomorphic encryption enables computations to be performed directly on encrypted data, without the need for decryption. This has significant implications for privacy-preserving technologies, as it allows for secure outsourcing of computation to untrusted servers, while maintaining the confidentiality of the data. Moreover, RSA, AES, and DES can be used for other aspects of security, such as key management and message authentication.

Unlike other review papers [22] in the literature, which typically provide a high-level overview of different FHE schemes, this paper goes a step further by providing a simple and step-by-step algorithm for each of the FHE schemes discussed. This makes the FHE schemes more accessible to a wider audience, including those who may not have a deep background in cryptography or computer science. By breaking down the algorithms into simple steps, readers can follow along and understand how the scheme works at a more fundamental level.

In addition to providing a simple and step-by-step algorithm for each FHE scheme, this paper also presents a comprehensive comparison of different FHE schemes and open-source libraries in a tabular form in Table 1 and Table 2 respectively. The table includes important features and characteristics of each FHE scheme, such as security assumptions, key sizes, supported operations, computational complexity, and limitations. For example, if a user needs an FHE scheme that can perform addition and multiplication operations, they can easily filter the table to identify which schemes meet these criteria. Additionally, if a user is concerned about the computational complexity of the FHE scheme, they can compare the performance metrics of different schemes side-by-side.

The rest of the article is structured in the following way. Section 2. explains the mathematical concept behind Homomorphism, followed by its classification and applications. Section 3 analyses the various FHE Schemes. Section 4 presents five open-source libraries implemented both in Python and C++.

## 2. Background

This section provides mathematical concepts and background of homomorphic encryption.

## 2.1. Homomorphism

Let us take an example to understand this better. Let us take two messages $m_1$ and $m_2$ and their encrypted cipher texts $c_1 = E(m_1)$ and $c_1 = E(m_1)$. If function E is homomorphic, then one can obtain the value of $E(m_1 + m_2)$ by using $c_1$ and $c_1$ without knowing the values of $m_1$ and $m_2$.

$$E(m_1 + m_2) = c_1 + c_2 = E(m_1) + E(m_2)$$

Imagine the above scenario with any operation "⋆", then we can define an Encryption Scheme ($E$) as homomorphic if it supports the following equation:

$$E(m_1 \star m_2) = E(m_1) \star E(m_2), \forall m_1, m_2 \in M$$

where $M$ is the set of all possible messages.

In abstract algebra, a structure-preserving map between two algebraic structures or groups is homomorphism. Let us take a set $S$ and an operation "⋆", that combines any two elements a and b to form another element, denoted $a \star b$ and $a, b \in S$. The qualifications for a set and operation $(S,\star)$ to be a group are as follows:

- Closure property:
  For all $a, b \in S$, the result of $a \star b \in S$.
- Associativity property:
  For all $a, b, c \in S$, $(a \star b) \star c = a \star (b \star c)$.
- Identity element:
  For an element $e \in S$, the equality $e \star a = a \star e = a$ hold. Here $e$ is the identity element of set $S$.
- Inverse element:
  For an identity element $e \in S$ and elements $a, b \in S$, $a \star b = b \star a = e$ holds.

Note:

- The identity element $e \in S$ is often taken as 1.
- The result of operation may differ if the order of operand is changed. For example, $a \star b \neq b \star a, \forall a, b \in S$.

A group homomorphism from group $(G,\star)$ to group $(H,\star)$ is defined as $f: G \to H$ and holds if

$$f(g \star g`) = f(g) \star f(g`), \forall g, g` \in G$$

Group homomorphism comes into play when testing if an encryption scheme is homomorphic or not. Assume an encryption scheme $(P, C, K, E, D)$ where

- $P$ = Plain text
- $C$ = Cipher text
- $K$ = Key
- $E$ = Encryption algorithm
- $D$ = Decryption algorithm

and $(P,\star)$ and $(C,\star)$ are groups of plain texts and cipher texts, respectively. The encryption algorithm maps from plain text group $(P)$ to cipher text group $(C)$ using k from Key $(K)$ is homomorphic for any operation "$\star$" if

$$E_k(a) \star E_k(b) = E_k(a \star b), \forall a, b \in P \,\&\, k \in K$$

Here, $k$ can be a symmetric key or a public key, depending on the encryption algorithm used.

Using the above equation, let us prove that RSA is homomorphic for "•" i.e. Modular multiplication. The plain text group and cipher text group, respectively, are $(P,\bullet)$ and $(C,\bullet)$. For any two plain texts $p_1, p_2 \in P$, and public key $k = (n, e)$,

$$E(p_1, k) = p_1^e (mod\ n)$$

$$E(p_2, k) = p_2^e (mod\ n)$$

$$E(p_1, k) \bullet E(p_2, k) = p_1^e \bullet p_2^e (mod\ n) = (p_1 \bullet p_2)^e (mod\ n) = E(p_1 \bullet p_2, k)$$

Therefore, RSA is homomorphic for modular multiplication operation (•).

## 2.2. Classification of Homomorphic Algorithms

There are limitations to homomorphic algorithms. The existing encryption schemes may not satisfy homomorphism for all kinds of operations and any number of operations. Some encryption algorithms are homomorphic for addition and multiplication operations only; some are homomorphic for an infinite number of subsequent operations and for just one multiplication operation, etc. Hence, they are classified into three homomorphic encryption schemes:

### 2.2.1. Partially Homomorphic Encryption (PHE)

PHE provides for encrypted data calculations, but only for a limited number of operations, often addition and multiplication. PHE is less computationally costly than other types of homomorphic encryption, but its utility is limited. Some examples include RSA [3], Goldwasser-Micali [4], El-Gamal [5], Benaloh [6], Paillier [7], Okamoto-Uchiyama [8], etc.

### 2.2.2. Somewhat Homomorphic Encryption (SWHE)

SWHE enables more complicated operations on encrypted data, such as exponentiation and polynomial evaluation. SWHE is more computationally demanding than PHE and offers more capabilities. Some examples include BGN [9], Polly Cracker scheme [10], etc.

### 2.2.3. Fully Homomorphic Encryption (FHE)

FHE supports arbitrary calculations on encrypted data, such as conditional operations, branching, and looping. FHE is the most computationally costly type of homomorphic encryption, but it also offers the most functionality [23]. These algorithms mostly make use of techniques like bootstrapping to maintain homomorphism. Some examples include Ideal Lattice-Based [2], FHE Schemes over Integers [11], LWE-Based [12], NTRU-Like [13], etc.

## 2.3. Applications

All three schemes of homomorphic encryption i.e. PHE, SWHE and FHE has made themselves useful in any field which deals with data processing. It can be utilized in outsourcing storage and computations sector where the custom can share data with the outsourcing corporation without disclosing its sensitive data to the company, while also allowing the companies to perform operations on the data.

[1] offers a system architecture capable of performing biometric identification in the encrypted domain, as well as gives and analyses an implementation based on two current homomorphic encryption techniques. It also examines the technological aspects and challenges in this environment.

Let's a sample client-server interaction scenario, the client needs to send some sensitive data to the server, and the server returns the data after performing some operations on the data. This can be achieved with or without using HE. Both the methods are demonstrated below:

### 2.3.1. Without Homomorphic Encryption

Client (C) has an asymmetric key pair $(pu_C, pr_C)$ and message $M$ that must be sent to server. Similarly, server (S) has its asymmetric key pair $(pu_S, pr_S)$ and function $f$ which will be applied on client message. To maintain confidentiality, client encrypts M using server's public key to form $E(M, pu_S)$ which is then sent to the server. The server decrypts $E(M, pu_S)$ using its private key $pr_S$ to get $M$ and performs the function $f$ on $M$ to get $f(M)$. $f(M)$ is encrypted using client's public key $pu_C$ before being sent to the client. The client receives $E(f(M), pu_C)$ and decrypts it using its private key to get $f(M)$.

In this scenario, since the server can see the message, it may pose a huge security threat to the client. When dealing with sensitive data, there should be way to prevent the server from viewing the raw sensitive data.

### 2.3.2. With Homomorphic Encryption

Client (C) has a homomorphic encryption function $He$, it's corresponding decryption function $Hd$ and a message $M$ that must be processed by the server. The server (S) has the function $f$ to be applied on client's data. Client encrypts the message $M$ using homomorphic encryption to get $He(M)$ and sends it to the server for computation. The server performs the operation $f$ homomorphically to get $He\big(f(M)\big)$ and sends it back to the client. The client decrypts the data using $Hd\Big(He\big(f(M)\big)\Big)$ to get $f(M)$.

Unlike 2.3.1., here the server performs its operation blindfolded since it cannot see the original message. The confidentiality of the client's message is intact, for both the public and server.

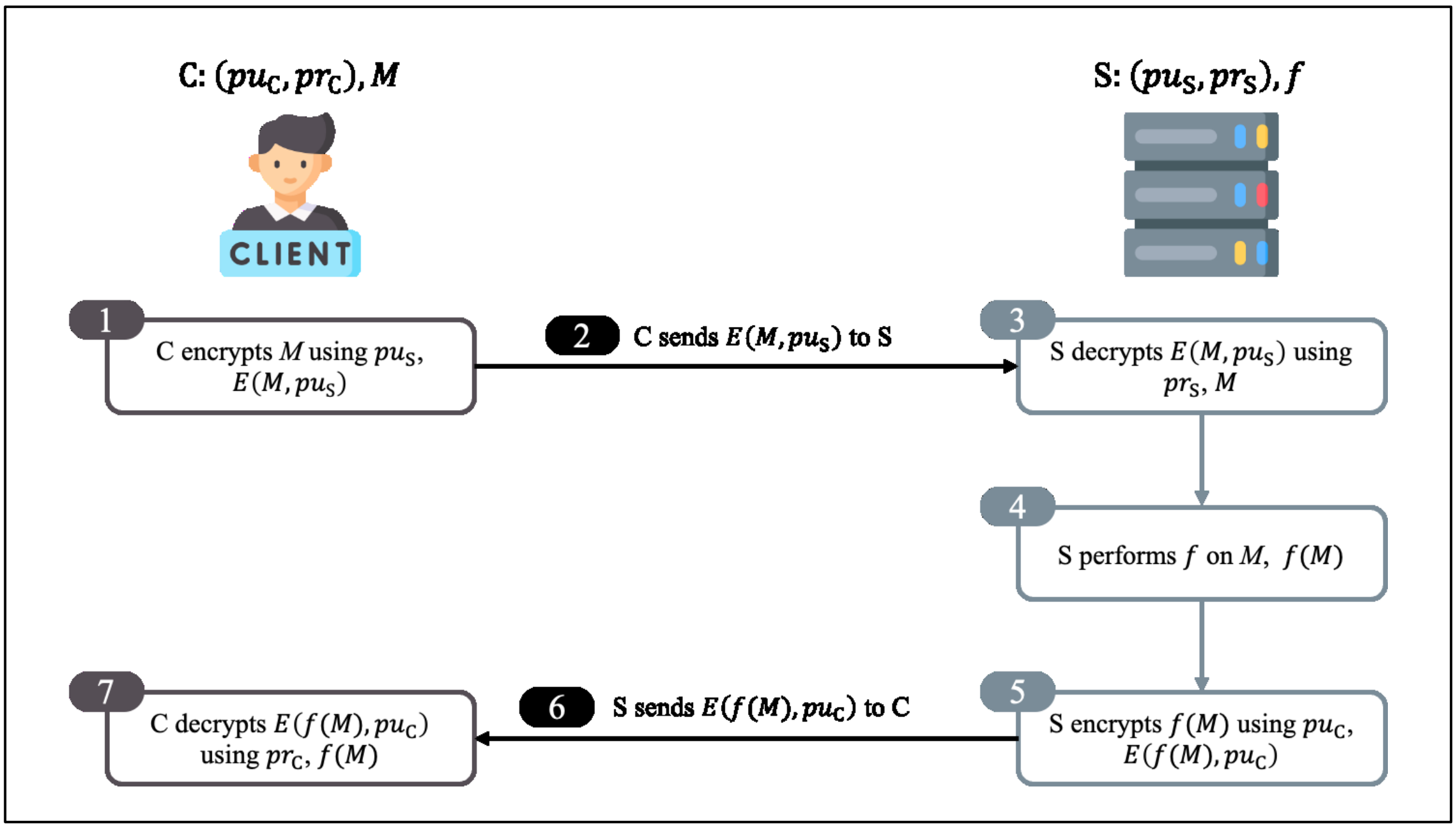

*Fig. 1. A client-server scenario without HE*

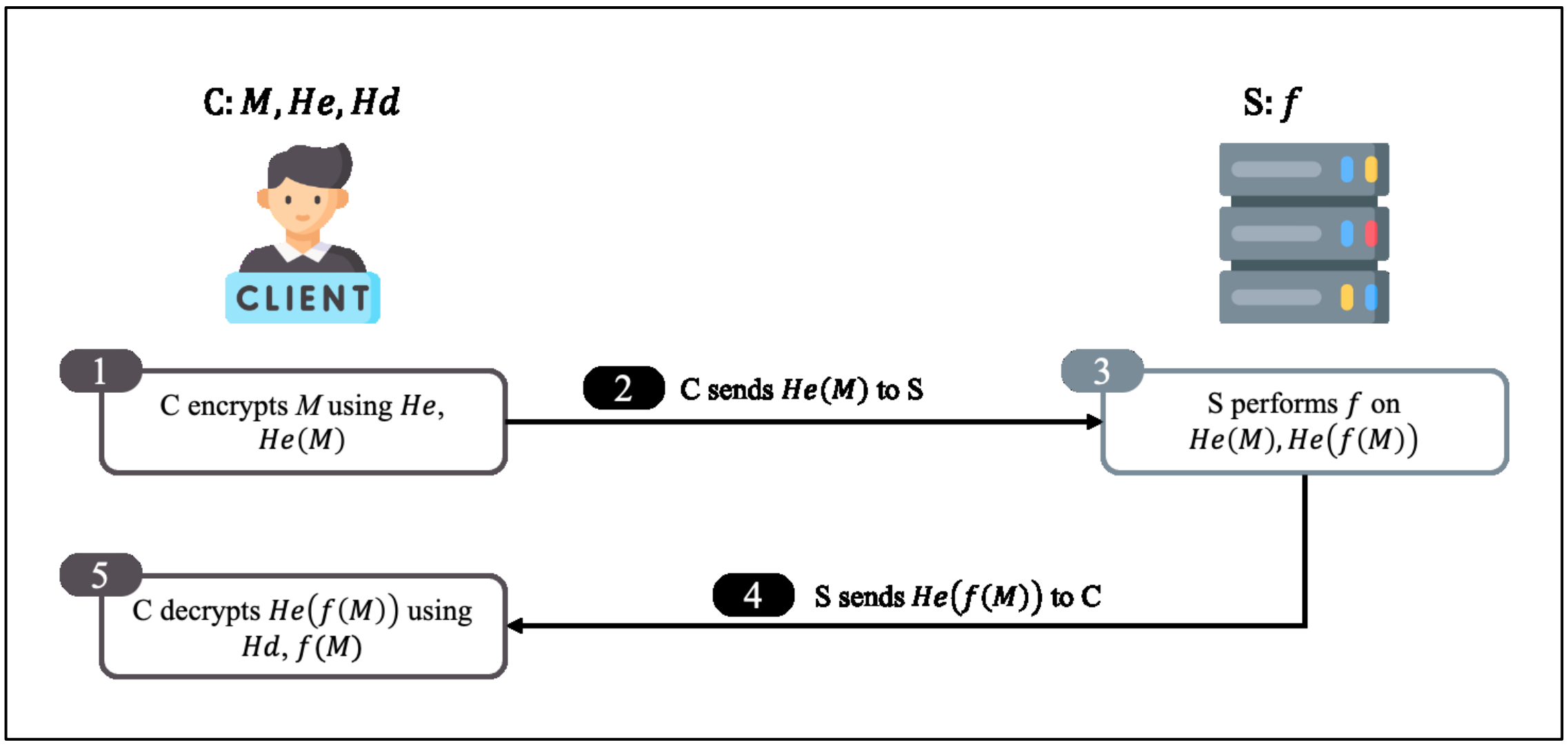

*Fig. 2. A client-server scenario with HE*

# 3. FULLY HOMOMORPHIC ENCRYPTION

To qualify as a FHE Scheme, an encryption algorithm should allow unlimited number of operations on the data while keeping the corresponding original data intact. The feasible concept of FHE was first published in 2009 by Gentry called Ideal lattice-based FHE scheme [2]. His work is promising, but computationally intensive. Ideal lattice-based FHE scheme was impractical for real-world scenario since the cost of operation was very high and was complex to implement.

Many new schemes and optimizations based on Gentry's work followed. They all had lattice-based approach in common which was hard to implement. Although this scheme was very promising, it was difficult to implement in real-world application due to high computational cost and difficult implementation dur to complex mathematics. Several optimizations and new schemes based on Gentry's work followed. To name a few FHE Schemes over Integers [11], LWE-Based [12], NTRU-Like [13], etc. We will discuss these schemes in detail below.

## 3.1. Ideal Lattice-based FHE Scheme

The first ever FHE Scheme was proposed in Gentry's thesis in 2009 and was based on GGH-type of encryption system [14]. Goldreich–Goldwasser–Halevi (GGH) lattice-based cryptosystem is an asymmetric cryptosystem based on lattices. The Goldreich-Goldwasser-Halevi (GGH) cryptosystem takes advantage of the fact that the nearest vector problem might be difficult.

A lattice $L$ with its basis $b_1, b_2, b_3, \ldots, b_n$ is formulated as follows:

$$L = \sum_{i=1}^{n} \overrightarrow{b_i} * v_i \, , v_i \in \mathbb{Z}$$

The basis of a lattice is not unique. There are infinitely many bases for a given lattice. A basis is called "good" if the basis vectors are almost orthogonal; otherwise, it is called "bad" basis of the lattice [16]. We know that the amount of noise in a cipher text shouldn't cross a threshold beyond which the plain text cannot be recovered. Therefore, from time-to-time the noisy cipher text must be processed to either eliminate or decrease the noise level. Ideal Lattice-based approach uses methods like squashing and bootstrapping, to reduce noise. This enables us to perform infinite number of operations on cipher text without losing the integrity of plain text.

Before understanding Ideal Lattice-based FHE Scheme by Gentry, you should understand the meaning of Ideal of Ring Theory. An ideal of a ring is a special subset of its elements. For example, even numbers for a set of Integers. Addition and subtraction of even numbers preserves evenness and multiplying an even number by any integer (even or odd) results in an even number; these closure and absorption properties are the defining properties of an ideal.

Gentry's SWHE scheme using ideals and rings is described below:

**Key Generation Algorithm:**

1. Choose two prime numbers $p$ and $q$ such that $q$ divides $p - 1$.
2. Choose a generator $g$ of the subgroup of $F_p *$ of order $q$.
3. Choose two elements $a, b$ uniformly at random from $F_q$ and compute $f = g^a * f^b$, where $f$ is a randomly chosen element of $F_p$.
4. The secret key is $(a, b)$ and the public key is $(h, p, q, g)$.

**Encryption:**

1. Let $m$ be the plaintext message.
2. Choose a small integer $t$ and $a$ random element $e$ in $F_p$.
3. Compute the ciphertext as $c = m * h^t * g^e \ mod \ p$.

**Decryption:**

1. Compute $s = at + be \ mod \ q$.
2. Compute $m' = c * g^{-s} \ mod \ p$.
3. Round $m'$ to the nearest integer to obtain the decrypted plaintext message.

**Homomorphic Addition:**

To add two ciphertexts c1 and c2, compute c1*c2 mod p.

**Homomorphic Multiplication:**

To multiply two ciphertexts $c_1$ and $c_2$, first compute $c' = c_1 * c_2 \ mod \ p$. Then, given a plaintext message $m$ with a small integer representation, compute $(c')^m \ mod \ p$ to obtain the ciphertext of the product.

The security of the SWHE scheme is based on the hardness of the learning with errors (LWE) problem, which involves the estimation of a random linear function of a noisy sample. The scheme provides limited homomorphic computations, such as addition and multiplication of encrypted messages, without revealing the underlying plaintext.

Gentry's FHE (Fully Homomorphic Encryption) approach employs a "squashing" strategy to deal with the noise that builds in ciphertexts after many homomorphic operations. The approach also contains a "bootstrapping" technique for refreshing the noise in ciphertexts, allowing for limitless homomorphic computations. Gentry's FHE system has the following procedure for squashing and bootstrapping.

**Squashing:**

1. Choose a small positive integer $L$.
2. Compute the function $f(x) = x^L - 1$.
3. Evaluate the function on the ciphertext $c$ to obtain a new ciphertext $c' = f(c) \ mod \ p$.
4. The squashing step reduces the noise in the ciphertext c by a factor of $L$, at the cost of losing information about the plaintext message. However, the information can be recovered using bootstrapping.

**Bootstrapping:**

1. Choose a fresh key pair $(a', s')$ using the same key generation algorithm as before.
2. Evaluate the decryption circuit on the squashed ciphertext $c'$ to obtain a new ciphertext $c''$ that encrypts the same plaintext message as $c$, but with noise reduced to a negligible level.
3. Compute a new public key $h' = (g, h'_1, \dots, h'_n)$ using the same method as before, where $h'_i = g^{a'i} * f^{s'i}$ for a randomly chosen $f$ in $F_p$.
4. Compute a new ciphertext $c'''$ that encrypts the same plaintext message as c, but with the new public key $h'$ and a much smaller level of noise.
5. The ciphertext $c'''$ can now be used as input for further homomorphic computations.

The bootstrapping step involves two decryption and encryption operations, and a new public key is needed for each bootstrapping operation. Therefore, bootstrapping is computationally expensive and limits the practicality of the FHE scheme for large-scale computations. However, it allows for unlimited homomorphic computations on encrypted data, making it a powerful tool for privacy-preserving data analysis and machine learning.

## 3.2. FHE Schemes over Integers

A new fully homomorphic encryption scheme was proposed in [11] that was based on the Approximate-Greatest Common Divisor (AGCD) problems. AGCD problems try to recover $p$ from the given set of $x_i = pq_i + r_i$ . The primary motivation behind the scheme is its conceptual simplicity. A symmetric version of the scheme is probably one of the simplest schemes.

The proposed symmetric SWHE scheme is described as follows:

**Key Generation Algorithm:**

1. Given a security parameter $\lambda$.
2. A random odd integer $p$ of bit length $\eta$ is generated. This will be treated as a private key.
3. Choose a random large number $q$.
4. Choose a small number $r$ such that $r \ll p$.

**Encryption Algorithm**

The message m ∈ {0, 1} is encrypted by using the following:

$$c = E(m) = m + 2r + pq$$

**Decryption Algorithm**

The following formula can be used for decryption:

$$m = D(c) = (c \bmod p) \bmod 2$$

**Homomorphism over Addition**

$$E(m_1) + E(m_2) = m_1 + 2r_1 + pq_1 + m_2 + 2r_2 + pq_2$$

$$= (m_1 + m_2) + 2(r_1 + r_2) + (q_1 + q_2)q$$

The output clearly falls within the ciphertext space and can be decrypted if the noise $|(m_1 + m_2) + 2(r_1 + r_2)| < {}^{p}/_{2}$. Since $r_1, r_2 << p$, a various number of additions can still be performed on ciphertext before noise exceeds ${}^{p}/_{2}$.

**Homomorphism over Multiplication**

$$E(m_1)E(m_2) = (m_1 + 2r_1 + pq_1)(m_2 + 2r_2 + pq_2)$$

$$= m_1m_2 + 2(m_1r_2 + m_2r_1 + 2r_1r_2) + kp$$

$N = m_1m_2 + 2(m_1r_2 + m_2r_1 + 2r_1r_2)$

The encrypted data can be decrypted if the noise is smaller than half of the private key. $N < {}^{p}/_{2}$. $N$ grows exponentially with the multiplication operation. This puts more restriction over the homomorphic multiplication operation than addition.

## 3.3. LWE-Based FHE Scheme

Brakerski and Vaikuntanathan's LWE-based fully homomorphic encryption (FHE) scheme [12] is a lattice-based cryptosystem that builds upon the Learning with Errors (LWE) problem. Here is a high-level description of the scheme:

**Key Generation**

1. Choose a prime number $p$ and an integer $q$ such that $q > p^2$.
2. Let $n$ be a positive integer and choose a random matrix $A \in Zq^{(n\times m)}$ and random vectors $s$, $e$, and $u \in Zq^n$.
3. Compute $b = (A, As + e)$ and the matrix $B = (b \mid u)$.
4. The public key is $(A, B)$, and the private key is kept secret.

**Encryption Algorithm**

To encrypt a binary message $m \in \{0,1\}$, generate a random vector $r \in Zq^n$ and a small noise vector $e' \in Zq^n$. Then, compute the ciphertext c as follows:

$$c = Ar + m * p + e'$$

The ciphertext $c$ is then sent to the recipient.

**Homomorphic Operations**

The scheme allows for homomorphic addition and multiplication of ciphertexts. Given two ciphertexts $c_1$ and $c_2$, the following operations can be performed:

- **Addition:** $c_1 + c_2 = c_1 + c_2$
- **Multiplication:** $c_1 * c_2 = (A * B')r + m_1 m_2 p + e_1 * B'r + e_2 Ar + e_1 e_2$

Here, $B'$ is the transpose of $B$, $m_1$ and $m_2$ are the plaintexts corresponding to $c_1$ and $c_2$, and $e_1$ and $e_2$ are the corresponding noise vectors.

**Decryption**

To decrypt a ciphertext $c$, compute the inner product of c with the private key vector $s$ modulo $p$, and then round to the nearest integer modulo 2:

$$m = round((c * s)/p) \, mod \, 2$$

This recovers the original message $m$.

Brakerski and Vaikuntanathan's FHE scheme improves upon Regev's scheme by allowing for more homomorphic operations before the noise in the ciphertexts grows too large. However, the scheme is still limited by the size of the noise in the ciphertexts, which ultimately limits the number of homomorphic operations that can be performed.

## 3.4. NTRU-Like FHE Scheme

NTRU (N-th degree TRUncated polynomial) is a public key cryptosystem based on the shortest vector problem (SVP) in a lattice. In recent years, NTRU has been used as a basis for constructing fully homomorphic encryption (FHE) schemes. The basic idea behind NTRU-Like FHE schemes is to use a variant of the NTRU public key cryptosystem as the underlying encryption scheme. The encryption process involves encoding the message into a polynomial and then adding noise to it. The decryption process involves recovering the message by finding the closest polynomial to the ciphertext polynomial in a certain norm.

**Key Generation:**

1. Choose integers $N$, $p$, and $q$, where $p$ and $q$ are large primes congruent to 1 modulo $2N$, and $p$ divides $q - 1$.
2. Generate a random polynomial $f(x)$ of degree $N - 1$ with coefficients in $\{-1, 0, 1\}$.
3. Compute the inverse polynomial $f^{-1}(x) \ mod \ q$.
4. Choose a small integer e and compute $g(x) = (1 + f(x))^e \ mod \ p$.
5. Public key is $(p, q, g(x))$ and private key is $(f(x), f^{-1}(x))$.

**Encryption:**

1. Encode the message m into a polynomial $m(x)$ of degree $N - 1$ with coefficients in $\{-1, 0, 1\}$.
2. Choose a small integer r and compute $h(x) = rg(x) + m(x) \ mod \ q$.
3. Send the ciphertext $(h(x))$.

**Decryption:**

1. Compute $c(x) = h(x) * f^{-1}(x) \ mod \ q$.
2. Compute $m(x) = round(c(x) \ mod \ p) \ mod \ 2$, where $round()$ is the function that rounds to the nearest integer.
3. The decrypted message is the polynomial $m(x)$.

**Homomorphic Addition:**

1. Given two ciphertexts $h_1(x)$ and $h_2(x)$, compute $h_3(x) = h_1(x) + h_2(x) \ mod \ q$.
2. Send the ciphertext $(h_3(x))$.

**Bootstrapping (Homomorphic Multiplication):**

1. Decrypt the ciphertext $h(x)$ to obtain $c(x) = h(x) * f^{-1}(x) \ mod \ q$.
2. Choose a random element a from $Z_p$.
3. Compute $c'(x) = (c(x) + a * f(x)) \ mod \ q$.
4. Compute $m'(x) = round(c'(x) \ mod \ p) \ mod \ 2$.
5. Compute $h'(x) = 2r * (g(x)^a) \ mod \ q$.
6. Compute $h''(x) = h(x) - h'(x) \ mod \ q$.
7. Compute $h3(x) = h''(x) + h'(x) * m'(x) \ mod \ q$.
8. The encrypted result is the ciphertext $(h3(x))$.

The above bootstrapping step can be repeated multiple times to enable deeper homomorphic circuits. The fully homomorphic aspect of these schemes comes from the fact that the NTRU cryptosystem is somewhat homomorphic. This means that the addition of two ciphertexts results in a ciphertext that can be decrypted to the sum of the corresponding plaintexts. Multiplication of ciphertexts, however, is not directly possible in the NTRU cryptosystem. Therefore, NTRU-Like FHE schemes use a technique called "bootstrapping" to perform homomorphic multiplication.

## 3.5. TFHE Scheme

TFHE is an open-source library for fully homomorphic encryption, distributed under the terms of the Apache 2.0 license [17]. The TFHE algorithm is a homomorphic encryption scheme that supports both addition and multiplication operations. It was proposed in 2018 and is designed to be more efficient than other FHE schemes, such as the BGV and FV schemes.

It also implements a dedicated Fast Fourier Transformation for the anticyclic ring $\mathbb{R}[X]/(X^N + 1)$, and uses AVX assembly vectorization instructions. The default parameter set achieves a 110-bit cryptographic security, based on ideal lattice assumptions. From the user point of view, the library can evaluate a net-list of binary gates homomorphically at a rate of about 50 gates per second per core, without decrypting its input. It suffices to provide the sequence of gates, as well as ciphertexts of the input bits. And the library computes ciphertexts of the output bits.

The key features of the TFHE algorithm include:

1. **Encryption:** The plaintext message is first encrypted using a symmetric key encryption scheme such as AES (Advanced Encryption Standard). This produces a ciphertext that is a stream of bits.
2. **Encoding:** The bits of the ciphertext are then encoded into a Torus, which is a mathematical structure that allows for efficient manipulation of the ciphertext using homomorphic operations.
3. **Homomorphic operations:** The TFHE algorithm supports both addition and multiplication operations on the encrypted data. These operations are performed using the encoded Torus values.
4. **Decryption:** The homomorphic result is then decoded back into a stream of bits and decrypted using the same symmetric key encryption scheme used in the encryption step.

The TFHE algorithm is designed to be efficient in terms of computational cost, memory usage, and ciphertext size. It achieves this by using a combination of symmetric key encryption and bitwise operations. Additionally, it supports rotation operations, which allows for efficient evaluation of circuits with loops or variable-length operations.

Table 1. FHE Schemes

| Scheme | Year | Security Level | Key Size | Homomorphic Operations | Implementation | Limitations |
|---|---|---|---|---|---|---|
| Gentry's FHE | 2009 | IND-CPA | 2^80 | Addition, Multiplication | Software | Slow evaluation speed |
| Brakerski-Gentry-Vaikuntanathan (BGV) | 2011 | IND-CPA | 2^80 | Addition, Multiplication, Rotation | Software, Hardware | Slow evaluation speed, large ciphertext expansion |
| Fan-Vercauteren (FV) | 2012 | IND-CPA | 2^40 | Addition, Multiplication, Rotation | Software, Hardware | Smaller ciphertext expansion, but slower |
| | | | | | | evaluation speed |
| Homomorphic Hashing | 2013 | IND-CPA | N/A | Hashing | Software | Limited to hash functions |
| Brakerski-Gentry 2 (BGV2) | 2014 | IND-CPA | 2^80 | Addition, Multiplication, Rotation | Software, Hardware | Smaller ciphertext expansion, but slower evaluation speed |
| Approximate Number Homomorphism (ANH) | 2015 | IND-CPA | N/A | Approximate addition, multiplication | Software | Limited accuracy |
| Gentry-Sahai-Waters (GSW) | 2016 | IND-CPA | 2^64 | Addition, Multiplication, Rotation | Software, Hardware | Smaller ciphertext expansion, but slower evaluation speed |
| TFHE | 2018 | IND-CPA | N/A | Addition, Multiplication, Rotation | Software, Hardware | Fast evaluation speed, but requires high computational power |
| Brakerski-Gentry-Levin (BGL) | 2018 | IND-CPA | 2^64 | Addition, Multiplication, Rotation | Software, Hardware | Smaller ciphertext expansion, but slower evaluation speed |

# 4. OPEN-SOURCE LIBRARIES

To make FHE more accessible and easier to use, several open-source libraries for FHE have been developed. These libraries provide a set of pre-built functions and APIs that allow developers to easily implement FHE in their applications without needing to understand the underlying mathematics and algorithms. Some of the popular open-source FHE libraries include HElib, SEAL, OpenFHE, TFHE, and HEAAN. These libraries support different FHE schemes and have varying levels of complexity and efficiency. Developers can choose the library that best suits their needs based on factors such as performance, ease of use, and compatibility with their existing systems.

Open-source FHE libraries are especially valuable for researchers and developers who are working on FHE-based applications but may not have the resources or expertise to develop their own implementation from scratch. By leveraging the work of others, they can quickly and easily incorporate FHE into their applications and advance the state of the art in secure computation.

## 4.1. HElib

HElib [18] is a C++ library for homomorphic encryption, supporting both the BGV and GSW schemes. It provides a simple and efficient API for performing homomorphic operations on encrypted data. HElib has been used in several real-world applications, including secure machine learning and privacy-preserving data analysis.

## 4.2. SEAL

SEAL (Simple Encrypted Arithmetic Library) [19] is a C++ library for homomorphic encryption, supporting the CKKS and BFV schemes. It provides a simple and user-friendly API for performing homomorphic operations on encrypted data. SEAL has been used in several real-world applications, including secure machine learning, privacy-preserving data analysis, and secure cloud computing.

## 4.3. OpenFHE

OpenFHE [20] is a C++ library for homomorphic encryption, supporting several schemes including BGV, BFV, CKKS, DM, and CGGI. It provides a flexible and modular API for performing homomorphic operations on encrypted data. Palisade has been used in several real-world applications, including secure machine learning, privacy-preserving data analysis, and secure cloud computing.

## 4.4. TFHE

TFHE (Fully Homomorphic Encryption over the Torus) [17] is a C++ library for homomorphic encryption, supporting the TFHE scheme. It is designed to be more efficient than other FHE schemes, such as the BGV and FV schemes. TFHE has been used in several real-world applications, including secure machine learning and privacy-preserving data analysis.

## 4.5. HEAAN

HEAAN (Homomorphic Encryption for Arithmetic of Approximate Numbers) [21] is a C++ library for homomorphic encryption, supporting the HEAAN scheme. It is designed for efficient computation on encrypted data using approximate numbers. HEAAN has been used in several real-world applications, including secure machine learning and privacy-preserving data analysis.

Table 2. Open-source libraries for fully homomorphic encryption (FHE)

| Library | Supported Schemes | Programming Language | API | License |
|---|---|---|---|---|
| HElib [18] | BGV, GSW | C++ | Simple and efficient | BSD 3-Clause |
| SEAL [19] | CKKS, BFV | C++ | Simple and user-friendly | MIT |
| OpenFHE [20] | BGV, BFV, CKKS, DM, CGGI | C++ | Flexible and modular | Apache 2.0 |
| TFHE [17] | TFHE | C++ | Efficient | LGPLv3 |
| HEAAN [21] | HEAAN | C++ | Efficient computation on approximate numbers | MIT |

# CONCLUSION

In conclusion, Fully Homomorphic Encryption (FHE) schemes allow computation to be performed directly on encrypted data, without requiring decryption. This is a powerful tool for preserving data privacy, as it enables secure outsourcing of computation to untrusted servers, while maintaining the confidentiality of the data. FHE schemes are based on a variety of mathematical problems, including lattice-based problems, AGCD problems, and NTRU-like problems. Each scheme has its own strengths and weaknesses, and the choice of scheme depends on the specific application and security requirements. Python libraries like Pyfhel and C++ libraries like OpenFHE provide an easy-to-use interface for implementing FHE schemes, enabling developers to experiment and prototype new applications. As FHE continues to advance, it has the potential to enable new privacy-preserving technologies and applications in fields such as healthcare, finance, and data analytics.